# The black hole-star formation connection over cosmic time

Chandra B. Singh[1], Nelitha Kulasiri[2], Max North[3], David Garofalo[2]

1. South-Western Institute for Astronomy Research, Yunnan University, University Town, Chenggong, Kunming 650500, People's Republic of China, email: chandrasingh@ynu.edu.cn
2. Department of Physics, Kennesaw State University, USA
3. Department of Information Systems, Kennesaw State University, USA

Abstract

Observations at low redshift have begun to tease out the star formation rate in active galaxies (AGN), which marks the beginning of the black hole-star formation connection over cosmic time. Star formation appears to depend on AGN type, cluster richness, and black hole accretion, but in ways that are not direct and have yet to be understood. Much of the confusion is that while some AGN appear to enhance star formation, others seem to suppress it. By implementing simplified, yet informed assumptions about AGN feedback on star formation, we show how AGN with jets might be dominated by two phases in which star formation is first enhanced, then suppressed. With this new element incorporated into our model, we make sense of radio and quasar mode behavior in the star formation rate – stellar mass (SFR-SM) plane for AGN. Due to jet feedback on star formation, jetted AGN tend to move upwards and rightward in the SFR-SM plane and then downward and to the right, past both the star-forming main sequence (SFMS) as well as the radio-quiet AGN. This picture allows us to predict the black hole connection to star formation as a function of the environment over the history of the universe.

1. Introduction

Given the dynamics, morphologies, and sizes of galaxies observed at late times in the universe, it appears inevitable that AGN feedback makes a difference in the cosmic evolution of galaxies. Without AGN feedback, for example, the masses of galaxies that result from numerical simulations are larger than observed (Silk & Rees 1998; Di Matteo et al. 2005; Silk & Mamon 2012). The black hole scaling relations (Magorrian et al. 1998; Gebhardt et al. 2000; Greene & Ho 2006) also strongly suggest that AGN feedback is needed to explain the dynamics of the gas, dust, and stars. Finally, star formation rates (SFR) in some AGN appears to be enhanced (Kalfountzou et al. 2012; Zinn et al. 2013), while in others it appears suppressed (Pawlik & Schaye 2009; Nesvadba et al. 2006, 2010; Morganti 2010; Breda et al. 2020). Either way, AGN appear to follow different tracks in the SFR-SM plane compared to the SFMS (Brinchmann et al. 2004; Elbaz et al. 2007; Speagle et al. 2014). Recent work on SFR in both radio-quiet and radio-loud AGN has been interpreted as evidence for negative AGN feedback on star formation (Comerford et al. 2020).

In this work, we extend our model for black hole accretion and jet formation to include AGN feedback on star formation by way of a simple coarse-grained yet informed assumption about jet-ISM interaction. Our picture prescribes radio-quiet AGN suppression of star formation while radio AGN on average experience phases in which star formation is enhanced, followed by longer phases in which it is suppressed. And the relative contribution of enhancement versus suppression depends on black hole mass and thus cluster richness. With these ideas in hand, we show how AGN evolve over cosmic time in the SFR-SM plane. In Section 2 we review the observational results we wish to interpret, describe our model, and apply them to the data. We then summarize and conclude.

2. Star formation in AGN

Comerford et al. (2020) have explored SFR and stellar mass for radio-loud and radio-quiet AGN, which we show in Figure 1, concluding that radio mode AGN suppresses star formation. We will argue that Figure 1 is evidence that radio mode AGN tends to first enhance star formation, then suppress it, and that black hole mass and therefore environment weighs the positive and negative feedback differently. Because our model singles out excitation level, the enhancement and suppression of star formation in radio mode versus quasar mode will allow us to connect star formation to excitation level so we will be able to explore star formation rates in high excitation radio galaxies (HERGs) and low excitation radio galaxies (LERGs). We show how the connection between jet morphology, black hole spin, excitation class, cluster richness, and feedback on star formation, emerges and will be able to explain the location of the radio mode AGN as well as their slope in Figure 1. In addition, we will make predictions about what such a figure would look like at higher redshift. In short, our model explains/predicts the interplay between AGN and star formation across cosmic time.

2.1 The AGN-star formation connection in the gap paradigm

To explain Figure 1, we enhance the gap paradigm for black hole accretion and jet formation (Garofalo, Evans & Sambruna 2010) to include jet feedback on star formation. Jetted AGN are a minority, which, in the model, is explained by associating FRII jets with counterrotation between accretion disk and black hole. Because most post-merger rotating black holes will be surrounded by corotating accretion disks, jets will be the minority (Garofalo, North, Belga, Waddell 2020). In Figures 2, 3, and 4, we add one additional column to previous figures that over the last decade have allowed us to explore many observations of AGN across space and time.

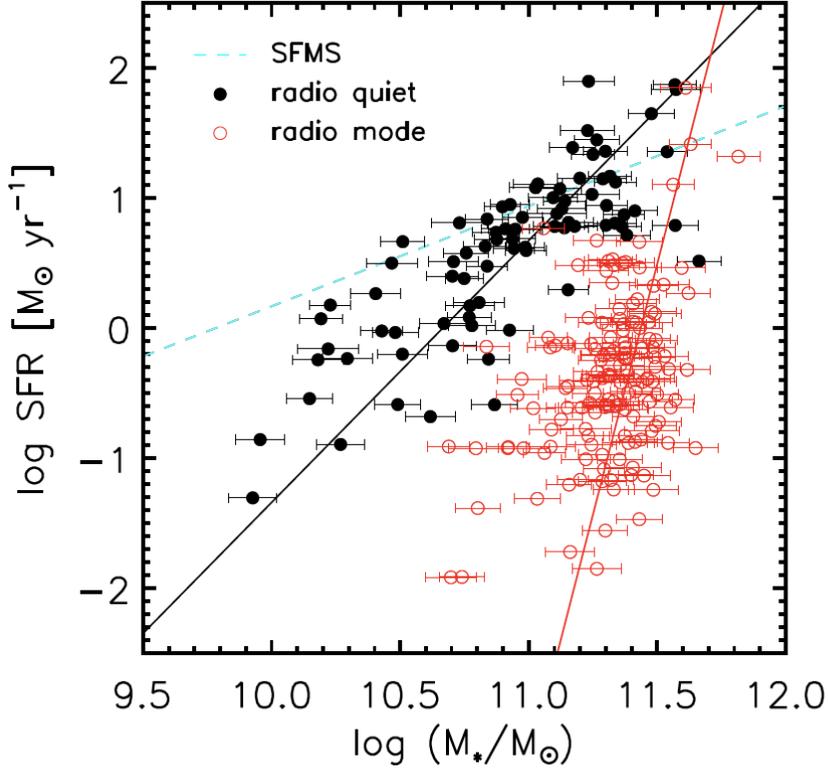

Figure 1: Radio quiet and radio mode AGN star formation versus stellar mass from Figure 3 of Comerford et al. (2020). Note how the "star forming main sequence"(SFMS) has the lowest slope. Despite the lower SFR among the radio mode AGN, the slope is larger than for the radio-quiet AGN. While the radio-quiet AGN show some degree of SFR suppression compared to the SFMS, we argue that the slope of the radio mode AGN is evidence for both SFR enhancement as well as SFR suppression.

Figures 2, 3, and 4 illustrate the time evolution of merger-triggered accretion onto rotating black holes in retrograde configuration. This is illustrated in the bottom panel of each figure with a negative '$a$' value. The merger leads to both star formation and AGN triggering but there is an offset between the two. Mihos & Hernquist (1996) estimate that between 65 and 85 % of the total supply of gas is converted into stars, on average, in 100 million years, give or take 50 million years. At some moment in time, we imagine that a radio quasar has been triggered. The effect of the jet does not act instantaneously on the ISM so we have no difference in terms of the ISM between a newly triggered radio quasar and a newly triggered radio quiet quasar. This means it is natural to start a radio quasar at the same place as a radio quiet quasar. This places it on the black radio quiet AGN line. While our purpose is to explore the evolution of star formation in radio mode AGN, our model also predicts the star formation behavior of radio quiet AGN. We limit this to emphasizing, with a black arrow, that radio quiet AGN experience star formation suppression, and as such, move downward and to the right in the SFR-SM plane.

Figure 2 describes the time evolution of merger-triggered accretion onto retrograde black holes in isolated environments where the black holes are on average the least massive although we are still dealing with masses above about $10^8$ solar masses. More precisely, the claim is not that all black holes in isolated environments have mass above $10^8$ solar masses; simply that the black holes in these environments that experience counterrotation, are the most massive (Garofalo, Christian & Jones 2019). Because of counterrotation, the gap region between the inner edge of the disk and the black hole is largest. These conditions are ideal for the greatest magnetic flux on the black hole (Reynolds et al. 2006; Garofalo 2009a) which gives rise to the largest Blandford-Znajek (BZ) and Blandford-Payne (BP) jets (Garofalo 2009b; Garofalo, Evans & Sambruna 2010; Connor et al. 2021). However, such conditions are not ideal for the radiative disk wind because the inner disk does not tap into the gravitational potential near the black hole. Hence, the radiative disk wind is relatively weak. Accretion imparts an inevitable (i.e. not model-dependent) time evolution involving spin-down of the black hole toward zero spin, followed by a spin up in the direction of the accretion material. Spinning the black hole down at the Eddington accretion limit requires about $8 \times 10^6$ years (Kim et al 2016). This timescale is fixed by the physics of accretion and is therefore neither ad-hoc nor fine-tuned. In the model, FRII jets are associated with retrograde accretion while FRI jets are associated with prograde accretion (details in Garofalo, Evans & Sambruna 2010 and references therein). The new element in these figures appears in the left column labeled "Star formation rate". We assume that FRII jets couple to the ISM in ways that enhance star formation (Kalfountzou et al. 2012; 2014). The degree of positive feedback on SFR depends on the length of time that SFR enhancement occurs as well as on the degree or strength of the feedback. The strength of the positive feedback scales with jet power, which scales with black hole mass in our model. The leftward column of Figure 2 indicates high rates of star formation, which drop naturally over time as the gas supply decreases. The radiative wind suppresses star formation, but this is initially weaker, as can be seen from the size of the red arrows in the Figure, due to the relatively large gap region between disk inner edge and black hole, and increases over time as the gap region decreases in size. We can make a simple quantitative connection between the optical spectrum of a star-formation-enhancing FRII, and the same object as it transitions through zero spin (i.e. away from its initial FRII phase). Thin disks around high spinning black holes in counterrotation are the least efficient because of the large gap regions. From standard thin disk theory, we can estimate the ratio of temperatures at various disk locations as a function of the innermost stable circular orbit ($r$SCO) which depends on spin. While temperature T and $r$SCO are related by the following,

$$T \sim r^{3/4} \, M^{1/4} \, [1-(r\text{SCO}/r)^{0.5}]^{1/4},$$

the relation between $r$SCO and the spin is approximately linear which we can parametrize as

$$r\text{SCO} \approx (6-4.2a) \, r_g$$

with *a*, whose range is -1 < *a* < 1 (negative represents counterrotation and positive corotation) and $r_g$ the radial distance in gravitational radii. Because of this, as the black hole spins down toward zero spin, the temperature at some radial location increases. At 10 gravitational radii from the event horizon, the temperature increase that results from a thin disk that evolves from high retrograde to zero spin, is about 10 percent. This is greater the closer one is to the inner disk edge. However, the inner disk edge moves inward. In general, if $T_i$ is the temperature at some radial location in the disk of a highly spinning black hole in counterrotation, and $T_f$ is the temperature at that same location when the spin has evolved over time, theory predicts that

$$T_f / T_i > 1.$$

This increase in disk efficiency with time is reflected in the disk spectrum, with the optical peak increasing over time, until the accretion rate drops sufficiently to compete with this. The lower optical peak associated with FRII star-formation-enhancement has been observed (Kalfountzou et al 2014).

Spinning the black hole up to high prograde values from the initial counterrotating state requires about $10^8$ years at the Eddington limit (Kim et al 2016). This is because the gas accreting onto the black hole from the inner disk edge has a quantifiable amount of angular momentum, which, if fed at the Eddington limit, results in a characteristic timescale. Like the time to spin the black hole down, this timescale is neither ad-hoc nor fine-tuned. The natural drop in accretion rates over time will, however, raise this to some hundreds of millions of years. Notice that the timescale for the prograde regime is at least an order of magnitude longer than for the retrograde one. At the most superficial level of analysis, this feature suggests that radio quasars are early stages in the AGN phenomenon and live comparably shorter lifetimes.

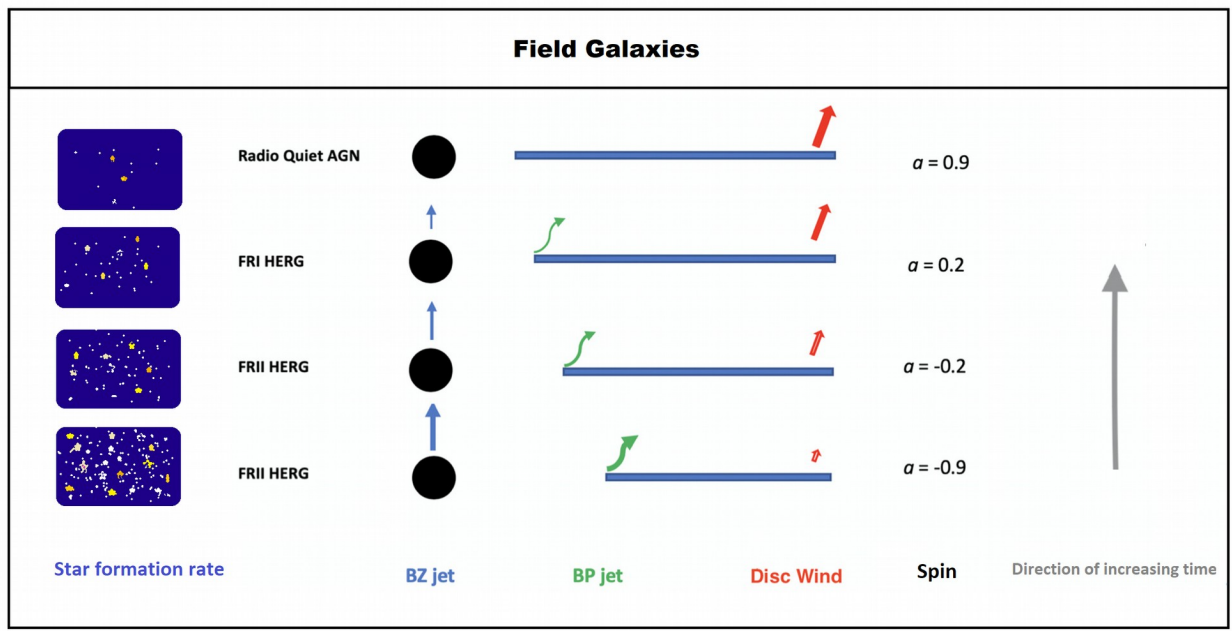

Figure 2: Time evolution of isolated AGN. For initially counterrotating black hole accretion

states, the initial positive feedback on star formation due to the FRII phase is followed by a brief negative jet feedback phase due to corotation and FRI jets. The disk wind increases in corotation, which also adds negative feedback on star formation. Because the black holes in these isolated environments are not very massive, this feedback is comparatively weaker and the star formation rate roughly follows the trend in the accretion rate, which simply drops over time. Because the disk is radiatively efficient and thin at all epochs, the label high excitation is associated with the presence of a jet in the term high excitation radio galaxy (HERG). When the jet is suppressed at a high enough prograde spin, the nomenclature changes. The left-hand column is meant as a simple visual so one can easily see the effect of FRIs on star formation. The arrow lengths and widths in the third and fourth columns capture the strength of the Blandford-Znajek and Blandford-Payne jets. The overall jet power results from combining both those jet mechanisms.

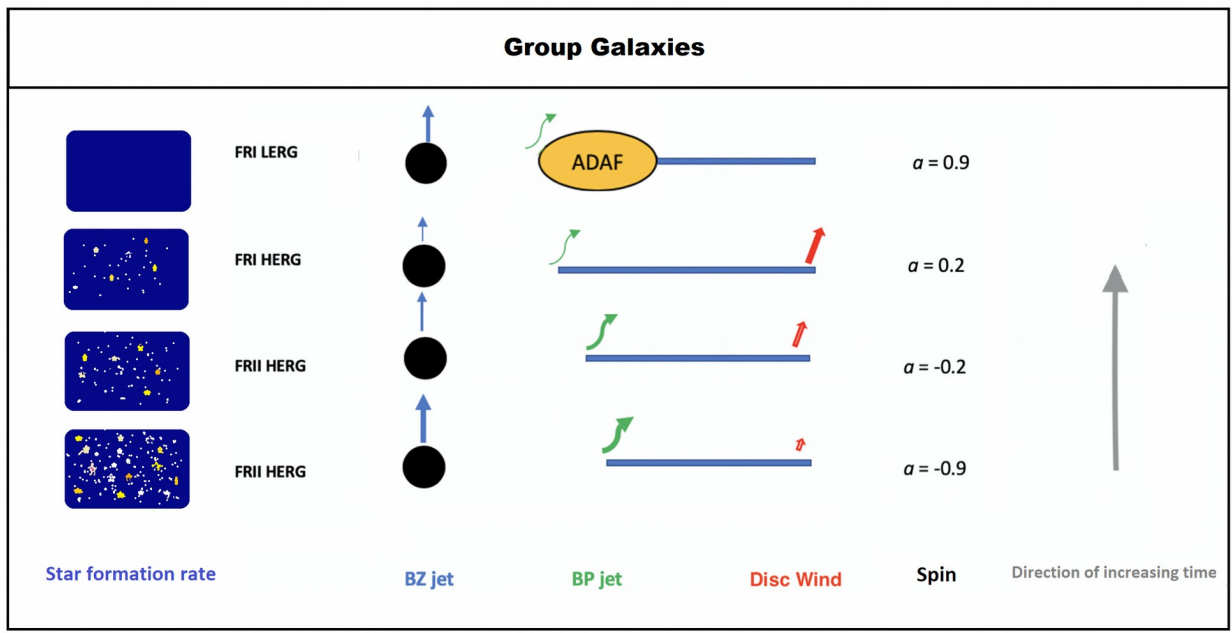

Figure 3: Time evolution of AGN in groups. For initially counterrotating black holes, the initial positive feedback on star formation due to the FRII phase is followed by a brief negative jet feedback phase due to corotation and FRI jets at late stages. The disk wind increases in corotation, which also adds negative feedback on star formation. Because the black holes in these environments are of greater average mass, this feedback is comparatively stronger than in field environments which enhances the negative feedback on star formation in corotation, making star formation drop more. But the black hole accretion rate also drops more compared to field galaxies.

Figure 3 describes the time evolution of merger-triggered accretion onto retrograde black holes in denser group environments where the black holes are on average more massive than in the field environments of Figure 2. Because of this, jet feedback is more effective in coupling to the ISM, and the form of accretion over a comparably long period changes as seen by the advection dominated accretion flow onto the black hole in the high spinning prograde configuration of the upper panel. Whatever amount of fuel remains to be accreted at this late stage, accretes at a rate that is at least two orders of magnitude less than at time t=0 when the system formed as a high spin retrograde system. Hence, in group environments, there is a prediction that FRI radio galaxies have high spins. This should contrast with cluster environments where similar objects exist but with intermediate spin values because it takes them a lot longer to evolve. While the FRII phase in this environment also enhances star formation as it does in field environments, the transition towards a late-stage ADAF with lower accretion rates will allow us to understand how such radio loud AGN will appear at lower SFR values but intermediate SM values in the SFR-SM plane (Figure 5).

Figure 4 describes the time evolution of merger-triggered accretion onto retrograde black holes in the densest, cluster environments, where the black holes are on average the most massive. Because of this, jet feedback is most effective in coupling to the ISM which changes the state of accretion more rapidly compared to groups. Because FRI jets suppress star formation, the black holes of Figure 4 are characterized by the strongest and most prolonged star formation suppression. The large suppression of star formation in FRI jet states is illustrated in the leftward column of Figure 4 with no stars. It is important to clarify that counterrotation is more likely to form in the densest environments not simply because the black holes tend to be the most massive. There is a connection with the larger range in the accretion rates onto the black holes in the mergers associated with such environments. The details behind stable counterrotating accretion is given by King et al (2005). While the relative angular momenta of the black hole and the disk are primary considerations, they can be translated into a condition for the relative masses of the black hole and the disk. Cluster environments offer the greatest range in cold gas mass and angular momenta, compared to less rich environments, and this translates into a greater fraction of systems with extreme values in the properties of the black hole and the disk, meaning disks with small angular momenta and mass and black holes with large mass (Garofalo, Christian & Jones 2019). While we do not go into further details here, this addresses the following question: Why don't all counterrotating black holes, regardless of environment, evolve or behave, in the same way? The answer is that rapidly spinning, very massive black holes, are more likely to be associated with massive cold gas flows onto them, in more isolated environments, which are more likely to form prograde accreting disks. The counterrotating black holes that do form in isolated environments, will tend to have less massive black holes, with less massive disks. As a result of this, the average counterrotating black hole in an isolated field environment will not behave like the average counterrotating black hole in a rich environment.

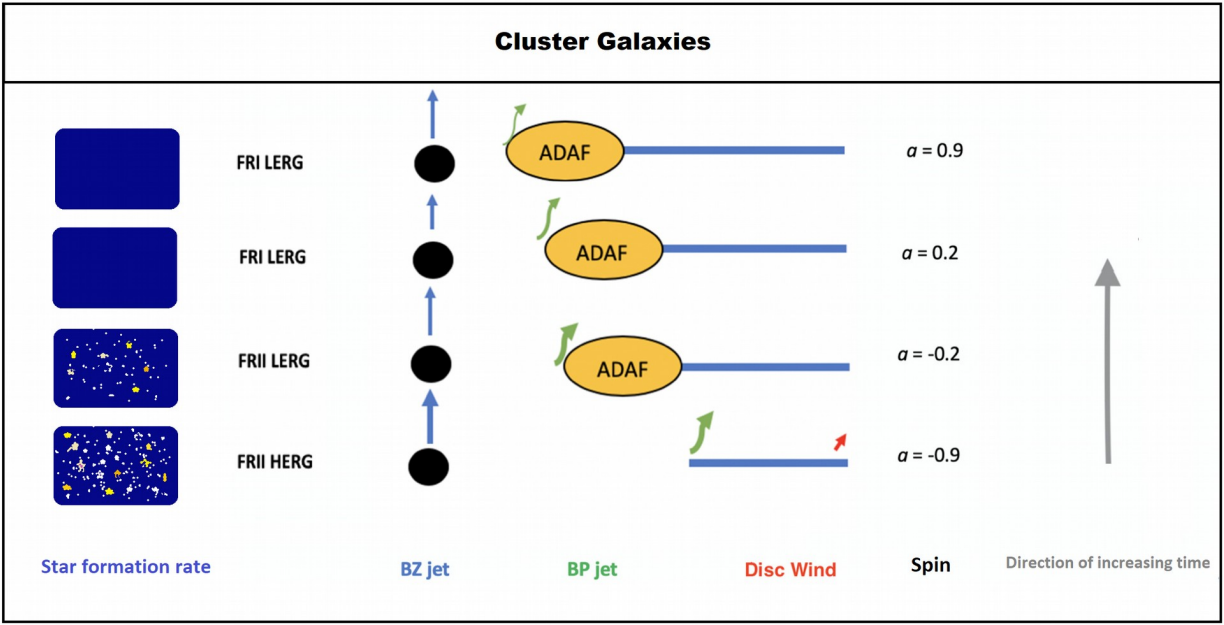

Figure 4: Time evolution of cluster AGN. For initially counterrotating black hole accretion, the initial positive feedback on star formation due to the FRII phase is followed by a long negative jet feedback phase due to corotation and FRI jets. Because the black holes in these rich environments are the most massive, this jet feedback is comparatively stronger than in other environments, which rapidly suppresses star formation.

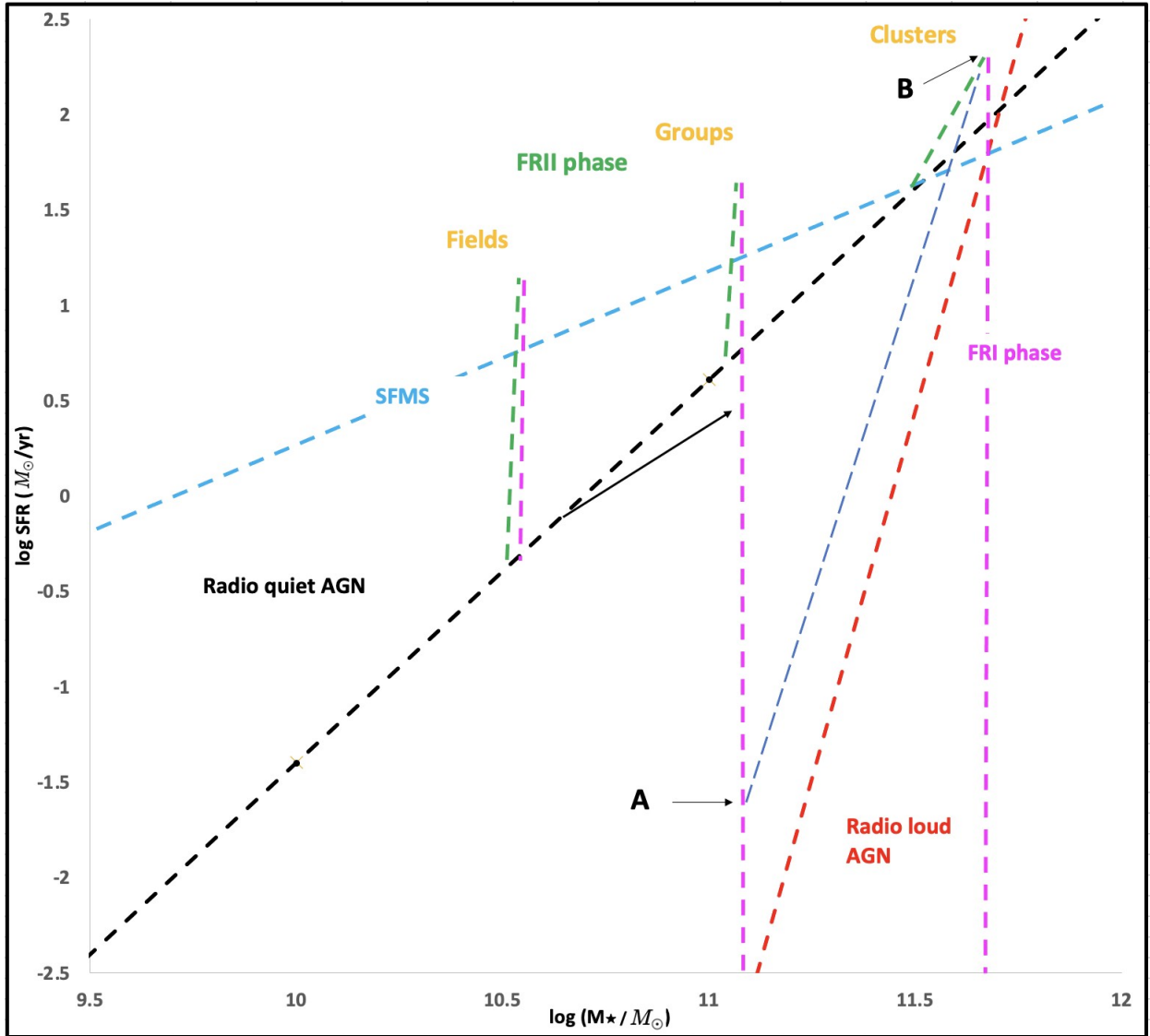

Figure 5: The AGN-SFR connection on the SFR-SM plane. The dashed black and dashed red lines are fits to the Comerford et al (2020) data. The parallel tracks illustrate varying degrees of SFR associated with different AGN phases. Post-merger-triggered AGN whose black holes end up in counterrotating configurations with respect to their accretion disks produce FRII jets that enhance star formation so such systems evolve upward but to the right as star formation increases (green curves). FRII phases are short-lived and followed by prograde accretion configurations which produce FRI jets. Such systems suppress star formation so the path experiences a drop in SFR while continuing to move rightward (pink curves). This constitutes the average cosmic evolution of radio mode AGN in the SFR-SM plane.

What Figures 2, 3, and 4 allow one to construct is the evolution of radio loud AGN in the SFR-SM plane which is shown in Figure 5. The black dashed line is taken from Comerford et al. (2020) and constitutes the fit to the radio quiet AGN data. Given a merger-triggered counterrotating black hole that triggers an FRII jet leading to enhanced star formation, the newly formed system transitions away from the SFR it would have as a radio-quiet AGN (such as those triggered in corotating accretion configurations) and moves along a path with a steeper slope (i.e. a green path). Notice how the black arrow diverges from the black dashed path. This indicates the slope of systems triggered in high spinning corotating configurations. This can be understood as negative feedback due to the radiative disk wind, which is larger for greater black hole spin. The different green paths branch from the black dashed path and are distinguished by the initial stellar masses which correspond to fields, groups, and clusters (from left to right). Considering the case of fields, we start with a SFR of about 0.3 M$_\odot$/year and a SM of $3.2 \times 10^{10}$ solar masses. This places it at log SFR = -0.5 and log SM = 10.5. The reason we choose values that place it on the radio quiet AGN line is due to the model prescription that if we remove the FRII jet, the system is a radio quiet AGN, as described previously. It is therefore natural to start each FRII phase (i.e. in fields, groups, and clusters) with a SFR that is equivalent to that of a radio quiet AGN. With respect to an equivalent radio quiet AGN, our AGN has an FRII jet which enhances the SFR by almost two orders of magnitude, peaking at log SFR = 1, which corresponds to an SFR of 10 M$_\odot$/year (Ichikawa et al 2021). Given that such a system accretes at the Eddington limit (bottom panel of Figure 2), its FRII phase will last less than $10^7$ years. If, for simplicity, we take the maximum SFR value during this FRII phase, we get

$$(10 \text{ M}_\odot/\text{year}) \times 10^7 \text{ years} = 10^8 \text{ M}_\odot$$

additional stars. This corresponds to a log SM = 10.506. In other words, the green path has a large slope on the SFR-SM plane. Once the FRII phase is completed, the radiative wind and eventually the FRI jet will suppress star formation. This is captured by the pink dashed line. Again, there is a relatively small shift in the horizontal value. These radio loud AGN in isolated environments remain on average in radiatively efficient disk states, which implies jet suppression at high black hole spin (upper panel of Figure 2). This means that after about $10^8$ years, the object becomes a radio quiet quasar or radio quiet AGN. Hence, we have an additional increase in the total number of stars that is no greater than

$$(10 \text{ M}_\odot/\text{year}) \times 10^8 \text{ years} = 10^9 \text{ M}_\odot.$$

Hence, we have log SM = log $(3.2 \times 10^{10} + 10^8 + 10^9)$ = 10.52, where we have added the additional 1 billion solar masses in stars to the total number of stars after SFR enhancement is complete. Because of jet suppression, however, these objects are no longer part of the radio loud class and become part of the radio quiet class. Hence, the pink dashed line ends when the conditions for jet suppression come into play. We have made a simple assumption here that the SFR is increased and then decreased by the same amount during the time prior to jet suppression. This need not be the case. Whether the pink dashed line should be longer or shorter is not part of physics we can determine, nor does it matter for field objects. What is key here is that field objects exit the radio loud class and star formation suppression continues to drop indefinitely in group and cluster environments. And this means that relative to the star formation suppression phase in groups and clusters, the pink dashed line will end at higher SFR for fields. This is captured in Figure 5 with FRII (green) SFR enhancement is followed by FRI (pink) SFR suppression in groups, fields, and clusters at increasing initial SM mass and SFR, respectively (yellow).

For completeness in our quantitative analysis, we adopt a maximum SFR of about 32 $M_\odot$/year on average for group environments (log SFR = 1.5) because the FRII jet feedback in Figure 3 is greater than in Figure 2. The FRII phases in groups tend to be about as long as FRII phases in fields, which adds no more than

$$(32 \; M_\odot/year) \times 10^7 \; years = 3.2 \times 10^8 \; M_\odot.$$

This means a change in our log SM from 11 to 11.01. Again, a steep slope for the FRII phase. In cluster environments, finally, FRII phases last on average about 100 times longer than in other environments due to ADAF accretion states entering early (second to bottom panel of Figure 4). Because the ADAF state has recently set in, we are close to the boundary between radiatively efficient and ADAF, which is $10^{-2}(dm/dt)_{Edd}$. SFR reaches a peak of 200 $M_\odot$/year which adds no more than

$$(200 \; M_\odot/year) \times 10^9 \; years = 2 \times 10^{11} \; M_\odot$$

in stars during the FRII phase. If the SM is $3.2 \times 10^{11} \; M_\odot$, we go from log ($3.2 \times 10^{11}$) = 11.5 to log ($3.2 \times 10^{11} + 2 \times 10^{11}$) = 11.72. FRI phases last billions of years in cluster environments, the longest of any environment because ADAF states come in early and last until no more gas is available (Figure 4). Hence, SFR continues to drop indefinitely. We use an average of 3 $M_\odot$/year over the next few billion years such that we add an additional

$$(3 \; M_\odot/year) \times 3.33 \times 10^9 \; years = 10^{10} \; M_\odot.$$

Hence, during the FRI SFR suppression phase (chosen to last 3.33 billion years), 10 billion stars are added and we go from log($3.2 \times 10^{11} + 2 \times 10^{11}$) = 11.72 to log($3.2 \times 10^{11} + 2 \times 10^{11} + 10^{10}$) = 11.724. Notice how the FRII ADAF phase of the second to bottom panel of Figure 4 is associated with the lowest slope on the SFR-SM plane (the green part of the cluster path). This rapid change in the state of accretion means that the cluster path from green to pink will take the longest timescale compared to groups and fields. This explains the three sets of FRII/FRI or SFR enhancement/suppression phases in fields, groups, and clusters. We now use these curves to explain and predict observations of radio loud AGN on the SFR-SM plane.

In field environments, radio mode AGN lasts on average no more than $10^8$ years. This is due to jet suppression at high black hole spin which is reached at the Eddington accretion limit in that time (Figure 2). As a result of this, the pink path shown in Figure 5 for field environments ends at a relatively high SFR compared to the pink paths in groups and clusters. This does not mean that SFR does not decrease further; simply that the object is no longer part of the radio mode AGN at lower SFR. At lower redshift, therefore, we expect to see little on the SFR-SM plane at relatively low SFR and low SM for radio mode AGN. In group environments, Figure 3 shows us that ADAF accretion phases come into play at late times (around 100 million years). This ensures absence of jet suppression (top panel of Figure 3) so the system remains in the radio loud AGN class as SFR continues to drop. We choose an arbitrary SFR of 0.03 $M_\odot$/year as the SFR that SFR suppression achieves on average in group environments after 1 billion years from the triggering of the original FRII HERG (lowest panel of Figure 3). This point is indicated as point A in Figure 5. This timescale is chosen because it corresponds to the time of transition between SFR enhancement and SFR suppression on average in cluster environments (labeled as point B in Figure 5). Point A and point B are connected by a blue dashed line to show how the radio mode AGN in cluster environments lag those in group environments as a result of the early instantiation of the ADAF phase in the former. We propose that the observed radio loud AGN red curve of Comerford et al (2020) results from the slow evolution in time of radio mode AGN in the densest environments compared to that in less dense environments.

It is important to notice that the alleged SFR enhancement phase predicted in our model and described in Figure 5 via the green dashed lines, implies that radio-loud AGN should be found also above the radio-quiet black dashed line. Yet, Figure 1 shows no radio-loud AGN above that line. Positive SFR feedback is relatively short-lived and therefore, in general, one expects fewer systems to be found above the radio-quiet AGN line than below it unless we catch the systems in their early stages, at higher redshift. In other words, once mergers become negligible and FRII HERG systems are no longer produced, 10 million years after the last merger, no such systems exist. In addition to that, all the FRII HERG, SFR enhancement phases, are followed by a transition through zero black hole spin. Hence, jets turn off during this transition and the objects exit the radio mode class. For initially retrograde accreting black holes in field and group environments, this transition will occur in radiatively efficient states, some of which lend themselves to the X-shaped radio galaxy phenomenon (Garofalo, Joshi, Yang et al. 2020). But many of these transition objects will cross the zero black hole spin regime at accretion rates that satisfy the following:

$$0.01(dm/dt)_{Edd} < dm/dt < (dm/dt)_{Edd}$$

In other words, the accretion rates are low enough that the FRII jet disappears while the system has not reached the prograde spin value of 0.1 for the compact jet (and hence to enter the FR0 classification) to reappear. In other words, some of the objects classified as radio-quiet AGN that are above the radio-quiet AGN line, may be transition objects that have completed their FRII phase and are transitioning towards the prograde jet and an ADAF phase as accretion rates continue to drop. We can estimate the time to transition from zero spin to a spin of 0.1, for example, in the prograde regime at dm/dt = (dm/dt)$_{Edd}$ and at dm/dt = 0.01(dm/dt)$_{Edd}$. The former requires a few million years while the latter requires 100 times that value, which brings it to a few hundred million years. Hence, such transition states oftentimes outlive FRII phases. The bottom line, however, is that as SFR-SM planes are constructed with higher redshift systems, more radio-loud AGN are predicted to populate the region above the radio-quiet AGN line but some subtlety needs to be taken into account given the evolution through zero black hole spin.

## 3. Conclusions

In this work, we have extended our model to include feedback on star formation by making simple yet informed assumptions about jets and radiative winds. We have shown that radio mode AGN tend to first enhance the SFR, which increases their slope in the SFR-SM plane, and then suppress it in a way that depends on black hole mass. Hence, the dependence translates into one that relates to cluster richness, with a slope for the observed radio mode AGN that will tend to be greater than that for radio quiet AGN. This difference in the slopes of the radio quiet and radio mode AGN has not been previously explored. We have motivated the idea that the bulk of the radio mode AGN shown in Figure 1 do not populate isolated environments, due to the fact that in such environments radio mode AGN have evolved out of their radio mode phase. The prediction of our model is that as surveys explore radio mode AGN at higher redshift, we will begin to populate the regions above the radio quiet mode AGN.

## 4. Acknowledgments

CBS is supported by the National Natural Science Foundation of China under grant no. 12073021. DG thanks J. Comerford for the data on our Figure 3.